\documentclass[a4paper,11pt,twoside]{article}
\usepackage[latin1]{inputenc}
\usepackage{amsfonts}
\usepackage{amsmath}
\usepackage{mathrsfs}
\usepackage{latexsym}
\usepackage{amssymb}
\usepackage{theorem}
\usepackage{epsfig}
\usepackage{graphicx}
\usepackage[dvips]{color}
\usepackage{array}
\usepackage{fancyhdr}

{\theorembodyfont{\itshape}\theoremheaderfont{\scshape}

}
{\theorembodyfont{\upshape} \theoremheaderfont{\scshape}

}

\newcommand{\fine}{\hfill $\Box$} 

\newenvironment{proof of}[1]{
\begin{trivlist}
\item[\hspace{\labelsep}{\sc\noindent\textsc{Proof of #1.}}]
}{\fine\end{trivlist}}

\def\Bernstein{{Bernshte\u\i n}}

\def\eps{\epsilon}

\def\beq{\begin{eqnarray}}
\def\eeq{\end{eqnarray}}
\def\beqn{\begin{eqnarray*}}
\def\eeqn{\end{eqnarray*}}

\font\bigsym=cmmi10 scaled \magstep3
   \def\smallprodi{\mathop{\lower 0pt\hbox{\bigsym\char'031}}}
\font\bbigsym=cmmi10 scaled \magstep4
   \def\prodi{\mathop{\lower 2pt\hbox{\bbigsym\char'031}}}
\font\bbbigsym=cmmi10 scaled \magstep5
   \def\Prodi{\mathop{\lower 2pt\hbox{\bbbigsym\char'031}}}

\begin{document}
\begin{sffamily}

\noindent \textmd{\LARGE{\bfseries{Bayesian Nonparametrics:
   Principles and Practice}}}

\bigskip
\noindent \textmd{\LARGE{\bfseries{Introduction}}}

\end{sffamily}
\bigskip

\thispagestyle{empty}

\noindent \textsc{Nils Lid Hjort, Chris Holmes,
   Peter M\"uller, and Stephen G.~Walker}



\bigskip
\noindent 
This extended preface is meant to explain why you are right 
to be curious about Bayesian nonparametrics -- why you
may actually need it and how you can manage to understand it
and use it. The preface also serves as an introductory chapter,
giving an overview of the aims and contents of the book. 
We also explain the background for how the book came into 
existence, delve briefly on the history of the still relatively
young field of Bayesian nonparametrics, and offer
some concluding remarks, pertaining to various
challenges and likely future developments of the area.


\setlength{\parskip}{2pt} 
\setlength{\parindent}{20pt}
\setlength\arraycolsep{2pt}

%

\section{Bayesian nonparametrics} 
\label{section:whatisit}

As modern statistics has developed over the past 
few decades various `dichotomies', where pairs
of approaches are somehow contrasted, are not as sharp 
as they appeared to be in the past. That some border 
lines appear more blurred than a generation or two ago 
is also seen regarding the contrasting pairs
`parametric vs.~nonparametric' and 
`frequentist vs.~Bayes'. It appears to follow
that `Bayesian nonparametrics' cannot be 
a very well-defined body of methods. 

\subsection{What is it all about?}

It is nevertheless an interesting exercise 
to delineate some of the implied regions 
of statistical methodology and practice 
by constructing a two-by-two table of sorts, 
via the two `factors' mentioned above; 
Bayesian nonparametrics would then be 
whatever is not found inside the other three categories. 

(i) `Frequentist parametrics' encompasses 
the core of classical statistics, involving 
methods associated primarily with maximum likelihood,
developed in the 1920ies and onwards. 
Such methods relate to various optimum tests, 
with calculation of p-values, optimal estimators, 
confidence intervals, multiple comparisons, 
etc. Some of the procedures stem from exact
probability calculations for models that are
sufficiently amenable to mathematical derivations,
while others relate to the application of 
large-sample techniques (central limit theorems,
delta methods, higher-order corrections involving
expansions or saddle-point approximations, etc.). 

(ii) `Bayesian parametrics' correspondingly
comprises classic methodology for prior and posterior
distributions in models with a finite (and often low) 
number of parameters. 
Such methods, starting from the premise that uncertainty
about model parameters somehow may be represented 
in terms of probability distributions, 
have arguably been in existence for more than a hundred years
(since the basic theorem that drives the machinery
simply says that the posterior density is proportional 
to the product of the prior density with the likelihood
function, which again relates to the Bayes theorem
of ca.~1763), 
but were naturally quite limited to a short list
of sufficiently simple statistical models and priors.
The applicability of Bayesian parametrics widened
significantly with the advent and availability 
of modern computers, say from ca.~1975 and onwards,
and then with the development of further numerical
methods and software packages pertaining to 
numerical integration and Markov chain Monte Carlo 
(MCMC) simulations, say from ca.~1990 and onwards. 

As for category (i) above,
asymptotics is often useful also for Bayesian parametrics, 
partly for giving practical and simple to use 
approximations to the exact posterior distributions
and partly for proving results of interest about
the performance of the methods, including aspects
of similarity between methods arising from 
frequentist and Bayesian perspectives. 
Specifically, frequentists and Bayesians agree 
in most matters, to the first order of approximation, 
for inference from parametric models, as the sample
size increases. The mathematical theorems that 
in various ways make such statements precise 
are sometimes collectively
referred to as `\Bernstein--von Mises theorems';
see e.g.~Le Cam and Young (1990, Ch.~7) 
for a brief treatment of this theme, including 
historical references going back not only to 
\Bernstein~(1917) and von Mises (1931)
but all the way back to Laplace (1810).
One such statement is that confidence intervals 
computed by the frequentist and the Bayesians 
(who frequently term them credibility intervals),
with the same level of confidence (or credibility),
become equal, to the first order of approximation,
with probability tending to one as the sample size increases.

(iii) `Frequentist nonparametrics' is a somewhat 
mixed bag, covering various different areas of 
statistics. The term has historically been associated 
with various test procedures that are or asymptotically 
become `distribution free', leading also to 
nonparametric confidence intervals and bands, etc.; 
for methodology related to statistics based on ranks 
(cf.~Lehmann, 1975); 
then progressively with estimation of probability
densities, regression functions, link functions etc., 
without parametric assumptions; and also with
specific computational techniques such as the bootstrap.
Again, asymptotics plays an important role, 
both for developing fruitful approximations 
and for understanding and comparing properties
of performance. A good reference book for 
learning about several classes of these methods 
is Wasserman (2006).

(iv) What ostensibly appears to remain for our 
fourth category, then, that of `Bayesian nonparametrics', 
are models and methods characterised by 
(a) big parameter spaces (unknown density 
and regression functions, link and response functions, etc.)~and
(b) construction of probability measures over these spaces.
Typical examples include Bayesian set-ups for 
density estimation (in any dimension), 
nonparametric regression with a fixed error distribution, 
hazard rate and survival function estimation
for survival analysis, without or with covariates, etc.
The division between `small' and `moderate' and `big'
for parameter spaces is not meant to be very sharp, 
and the scale is interpreted flexibly 
(see e.g.~Green and Richardson, 2001, for some 
discussion of this). 

It is clear that category (iv), which is the focus of our book,
must meet challenges of a greater order than for
the other three categories. The mathematical 
complexities are more demanding, since placing
well-defined probability distributions 
on potentially infinite-dimensional spaces 
is inherently harder than for Eucledian spaces.
Added to this is the challenge of `understanding 
the prior'; the ill-defined transformation from 
so-called `prior knowledge' to `prior distribution'
is hard enough for elicitation in lower dimensions
and of course becomes even more challenging
in bigger spaces. Furthermore, the resulting algorithms, 
e.g.~for simulating unknown curves or surfaces 
from complicated posterior distributions, 
tend to be more difficult to set up and to test properly.

Finally, in this short list of important subtopics,
we must note that the bigger world of nonparametric Bayes 
holds more surprises and occasionally exhibit 
more disturbing features than what the smaller 
and more comfortable world of parametric Bayes does. 
It is a truth universally acknowledged
that a statistician in possession of 
an infinity of data points must be in want of the truth -- but
some nonparametric Bayes constructions actually
lead to inconsistent estimation procedures, 
where the truth is not properly uncovered when
the data collection grows. Also, the \Bernstein--von Mises
theorems alluded to above, which hold very generally 
for parametric Bayes problems, tend not to hold as easily 
and broadly in the infinite-dimensional cases. 
There are e.g.~important problems where the nonparametric 
Bayes methods obey consistency (the posterior distribution 
properly accumulates its mass around the true model,
with increased sample size), but with a different 
rate of convergence than that of the natural frequentist
method for the same problem. Thus separate classes
of situations typically need separate scrutiny,
as opposed to theories and theorems that apply 
very grandly. 

It seems to us to be clear that the potential list
of good, worthwhile nonparametric Bayes procedures
must be rather longer, so to speak, than the 
already enormously long lists of Bayes methods
for parametric models, simply because bigger spaces
contain more than smaller ones. A book on 
Bayesian nonparametrics must therefore limit itself
to some of these worthwhile procedures. 
A similar comment applies to {\it the study} 
of these methods, in terms of performance,
comparisons with results from other approaches, 
etc.~(making the distinction between the 
construction of a method and the study 
of its performance characteristics). 

\subsection{Who needs it?}

Most modern statisticians have become well acquainted 
with various non- and semiparametric tools, on the one hand 
(nonparametric regression, smoothing methods, 
classification and pattern recognition, 
proportional hazards regression, copulae models, etc.), 
and with the most important simulation tools, 
on the other (rejection-acceptance methods, 
MCMC strategies like the Gibbs Sampler 
and the Metropolis algorithm, etc.),
particularly in the realm of Bayesian applications,
where the task of drawing simulated realisations
from the posterior distribution is the main operational job.
The {\it combination} of these methods 
is becoming increasingly popular and important
(in a growing number of ways), and each such 
combination may be said to carry the stamp 
of Bayesian nonparametrics. 

To answer the question of why combining nonparametrics
with Bayesian posterior simulations is becoming
more important, one component is related to 
practical feasibility, in terms of software packages
and implementation of algorithms. The other component
is that such solutions contribute to the solving
of actual problems, in a steadily increasing range
of applications, as indicated in this book, and
as seen at workshops and conferences dealing with
Bayesian nonparametrics. The steady influx of good
real-world application areas contributes both
to the sharpening of tools and to the sociological
fact that not only hard-core and classically oriented 
statisticians, but also various schools of other researchers 
in quantitative disciplines, lend their hands to work 
in variations of nonparametric Bayes methods. 
Bayesian nonparametrics is used by researchers working
in finance, geosciences, botanics, biology, epidemiology, 
forestry, paleontology, computer science, machine learning, 
recommender systems, etc.

By pre-fixing various methods and statements 
by the word `Bayesian' we are already acknowledging
that there are different schools of thought in
statistics -- Bayesians place prior distributions
over their parameter spaces while parameters are
fixed unknowns for the frequentists. 
We should also realise that there are different 
trends of thought regarding how statistical methods 
are actually used (as partly opposed to how they 
are constructed). In an engaging discussion paper, 
Breiman (2001) argues that contemporary statistics 
lives with a snowean `two cultures' problem. 
In some applications the careful study and 
interpretation of finer aspects of the model 
matter and are of primary concern, 
as in various substantive sciences -- an ecologist
or a climate researcher may place great emphasis
on a finding that a certain statistical 
coefficient parameter is positive, for example,
as this might be tied to scientifically relevant
questions of identifying whether a certain
background factor really influences a phenomenon
under study. 
In other applications such finer distinctions
are largely irrelevant, as the primary goals
of the methods are to make efficient predictions
and classifications of a sufficient quality.
This pragmatic viewpoint, of making good enough
`black boxes' without specific regard to the
components of the box in question, is valid
in many situations -- one might be satisfied
with a model that predicts climate parameters 
and the number of lynx in the forest, without 
always needing or aiming to understand the finer 
mechanisms involved in these phenomena.

This continuing debate is destined to play a role
also for Bayesian nonparametrics, and the right answer
to what is more appropriate, and to what is more
important, would be largely context-driven. 
A statistician applying Bayesian nonparametrics
may use one type of model for uncovering
effects and another for making predictions
or classifications, even when dealing with 
the same data. Using different models for different
purposes, even with the very same data set, 
is not a contradiction in terms, and relates
to different loss functions and to themes
of interest driven inference; cf.~various 
focussed information criteria for model selection
(see Claeskens and Hjort, 2008, Ch.~6).

It is also empirically true that some statistics 
problems are easier to attack using Bayesian methods,
with machineries available that make analysis 
and inference possible, in the partial absence
of frequentist methods. This picture may of course
be shifting with time, as better and more refined
frequentist methods may be developed also 
for e.g.~complex hierarchical models, but the
observation reminds us that there is a necessary
element of pragmatism in modern statistics work; 
one uses what one has, rather than spending 
three more months for developing alternative methods.
An eclectic view of Bayesian methods, also among
those statisticians hesitant to accept all of
the underlying philosophy, is to nevertheless use them,
as they are practical and have good performance.
Indeed a broad research direction is concerned
with reaching performance related results about
classes of nonparametric Bayesian methods,
as partly distinct from the construction of the
models and methods themselves 
(cf.~Chapter 2 in this book and its references).
For some areas in statistics, then, including 
some surveyed in this book, there is
an `advantage Bayes' situation. A useful reminder
in this regard is the view expressed 
e.g.~by Art Dempster (see Wasserman, 2008):
`a person cannot be Bayesian or frequentist;
rather, a particular {\it analysis} can be Bayesian of frequentist'.
Another and perhaps humbling reminder is 
Good's (1959) lower bound for the number 
of different Bayesians (46,656, actually),
a bound that perhaps needs to be revised upwards
when discussion concerns nonparametric Bayesians. 

\subsection{Why now?}

Themes of Bayesian nonparametrics have engaged 
statisticians for about forty years, but now,
as in around 2010 and onwards, the time is ripe 
for further rich developments and applications
of the field. This is due to a confluence of 
several different factors: 
the availability and convenience of computer programmes 
and accessible software packages, loaded down
to the laptops of the modern scientists,
along with methodology and machinery for finessing
and fine-tuning these algorithms for new applications; 
the increasing accessibility of statistical models
and associated methodological tools for taking
on new problems (leading also to the development
of further methods and algorithms); 
various developing application areas parallelling
statistics, that find use for these methods 
and sometimes develop them further; 
and the broadening meeting points for the 
two flowing rivers of nonparametrics (as such)
and Bayesian methods (as such).

Elements of the growing trend and importance 
of Bayesian nonparametrics can also be traced
in the archives of conferences and workshops
devoted to such themes. In addition to having
been on board in various broader conferences
over several decades, an identifiable subsequence 
of workshops and conferences specifically set up 
for Bayesian nonparametrics per se has
developed as follows, with a rapidly growing 
number of participants: 
Belgirate (1997),
Reading (1999),
Ann Arbor (2001),
Rome (2004),
Jeju (2006),
Cambridge (2007),
Turin (2009).
Monitoring the programmes of these conferences
one learns that development has been and remains 
steady, both regarding principles and practice.

Two more long-standing series of workshops are of interest 
to researchers and learners of nonparametric Bayesian statistics.
The BISP series (Bayesian inference for stochastic processes) 
is focussed on nonparametric Bayesian models related to 
stochastic processes. Its sequence up to the time
of writing reads 
Madrid (1998),
Varenna (2001),
La Mange (2003),
Varenna (2005),
Valencia (2007), 
Brixen (2009),
alternating between Spain and Italy. 
Another related research community is defined by 
the series of research meetings on Objective Bayes methodology.
The coordinates of the O'Bayes conference series history are 
Purdue, USA (1996), 
Valencia, Spain (1998), 
Ixtapa, Mexico (2000),
Granada, Spain (2002), 
Aussois, France (2003), 
Branson, USA (2005), 
Rome, Italy (2007),
Philadelphia, USA (2009). 

\section{The aims, purposes and contents of this book} 
\label{section:aims}

The present book has in a sense grown out of a certain
event. The book reflects this particular origin,
but is very much meant to stand solidly and
independently on its constructed feet,  
as a broad text on modern Bayesian nonparametrics;
in other words, readers do not need to know about
or take into account the event that led to 
the book being written. 

\subsection{A background event}

The event in question was a four-week programme 
on Bayesian nonparametrics hosted by the 
Isaac Newton Institute of Mathematical Sciences 
at Cambridge, UK, in August 2007, and organised by the four authors. 
In addition to involving a core group of some twenty 
researchers from various countries, the programme organised
a one-week international conference with about
a hundred participants. These represented an
interesting modern spectrum of researchers 
whose work in different ways is related to
Bayesian nonparametrics -- those engaged in 
methodological statistics work, from university departments
and elsewhere; statisticians involved
in collaborations with researchers from 
substantive areas (like medicine and biostatistics,
quantitative biology, mathematical geology, 
information sciences, paleontology); 
mathematicians; machine learning researchers; 
and computer scientists.

For the workshop, the organisers selected four 
experts to provide open tutorial type forum lectures, 
representing four broad, identifiable
themes pertaining to Bayesian nonparametrics. 
These were seen not merely as `four themes of interest',
but as closely associated with the core models,
the core methods, and the core application areas,
of nonparametric Bayes. These tutorials were 

-- Dirichlet processes, related priors and posterior asymptotics
(by S.~Ghosal); 

-- Models beyond the Dirichlet process 
(by A.~Lijoi, with I.~Pr\"unster as co-author);

-- Nonparametric Bayes applications to biostatistics
(by D.B.~Dunson); and

-- Bayesian nonparametrics in machine learning
(by Y.W.~Teh, with M.I.~Jordan as co-author). 

\noindent 
The programme and the workshop were evaluated 
(by the participants and other parties) as having been
very successful, by having bound together different
strands of work and perhaps by opening some doors
to further research work of promise, both 
theme-wise and person-wise. The experiences 
made it clear that nonparametric Bayes is 
an important growth area, with various 
side-streams that perhaps risk evolving 
too much by themselves if they do not make connections
with the core field or with other of its components.
All of this led to the idea of creating the present book. 

\subsection{What does this book do?}

We have chosen to structure the book around
these four core methods and core themes,
associated with the tutorials mentioned above,
and here appearing in the form of invited chapters. 
These are then complemented by `extension chapters', 
as follows:

-- Bayesian nonparametric methods: Motivation and ideas
(by S.G.~Walker, extending Ghosal's chapter);

-- Further models and applications
(by N.L.~Hjort, extending Lijoi and Pr\"unster's chapter); 

-- More nonparametric Bayesian models for biostatistics
(by P.~M\"uller and F.~Quintana, extending Dunson's chapter); and 

-- Bayesian nonparametrics for supervised and
unsupervised learning
(by J.~Griffin and C.~Holmes, extending Teh and Jordan's chapter). 

The extension chapters provide discussion, further 
developments, and links to related areas. 

As explained at the end of the previous section, 
it would not be possible to have `everything important'
inside a single book, in view of the size of the 
expanding topic. It is our hope and view, however,
that the dimensions we have probed are sound, deep 
and relevant ones, and that different strands 
of readers will benefit from working their way 
through some or all of these.

The {\it first} core theme (Chapters~1 and~2) 
is partly concerned with some of the cornerstone classes
of nonparametric priors, including the Dirichlet process
and some of its relatives. Mathematical properties are
investigated, including characterisations of the posterior
distribution. The theme also encompasses properties
of the behaviour of the implied posterior distributions,
and, specifically, consistency and rates of convergence.
Bayesian methodology is often presented as essentially 
a machinery for coming from the prior to the posterior
distributions, but is at its most powerful when 
coupled with decision theory and loss functions.
This is true for the nonparametric situations as well,
as also discussed inside this first theme. 

The {\it second} main theme (Chapters~3 and 4) 
is mainly occupied with the development of 
the more useful nonparametric classes of priors 
beyond those related to the Dirichlet processes
mentioned above: completely random measures, 
neutral to the right processes, the Beta process, 
partition functions, clustering processes, 
models for density estimation, stationary time
series with nonparametrically modelled covariance
functions, models for random shapes, etc., 
along with many application areas,
such as survival and event history analysis, 

The third and fourth core themes are more application
driven that the two first ones. 
The {\it third} core theme (Chapters~5 and 6) 
focusses on biostatistics. Topics discussed
and developed include personalised medicine 
(a growing trend in modern biomedicine), 
hierarchical modelling with Dirichlet processes, 
clustering strategies and partition models, 
and functional data analysis. 

Finally the {\it fourth} main theme (Chapters~7 and 8)
represents the important and growing application area
often referred to as machine learning. 
Hierarchical modelling, again with Dirichlet processes 
as building blocks, lead to algorithms 
that solve problems in information
retrieval, multi-population haplo-type phasing,
word segmentation, speaker diarisation, 
and so-called topic modelling. The models that 
help accomplishing these tasks include Chinese 
restaurant franchises and Indian buffet processes,
in addition to extensive use of Gaussian processes,
priors on function classes such as splines,
free-knot basis expansions, MARS and CART, etc.

\subsection{How to teach from this book}

Our book may be used as the basis for a Master 
or PhD level course in Bayesian nonparametrics.
Various options exist, for different audiences
and for different levels of mathematical skills.
One venue, for perhaps a typical audience 
of statistics students, is to concentrate 
on core themes two (Chapters~3 and~4) and three 
(Chapters~5 and 6), supplemented with computer exercises 
(drawing on methods exhibited in these chapters, 
and using e.g.~the software package described
in Jara, 2007). 
A course building upon the material in these chapters 
could be focussed on data analysis problems and 
typical data formats arising in biomedical research problems. 
Nonparametric Bayesian probability models would be 
introduced as and when needed to address the data 
analysis problems.
   
More mathematically advanced courses could then 
include more of core theme one (Chapters~1 and~2).
Such a course would be naturally more centred around 
a description of nonparametric Bayesian models
and include applications as examples to illustrate the models.
A third option is a course designed for an audience with interest
in machine learning, hierarchical modelling, etc. 
It would be focussed on core themes two (Chapters~2 and 3) 
and four (Chapters~7 and~8).

Natural prerequisites for such courses as briefly 
outlined here, and by association for working with this book, 
would include basic statistics courses 
(regression methods associated with generalised linear models,
density estimation, parametric Bayes),
perhaps some survival analysis (hazard rate
models, etc.), along with basis skills with
simulation methods (MCMC strategies). 

\section{A brief history of Bayesian nonparametrics} 
\label{section:history}

Lindley (1972) noted in his review of general Bayesian methodology 
that Bayesians up to then had been `embarrassingly silent' 
in the area of nonparametric statistics. He pointed
out that there were in principle no conceptual difficulties
with combining `Bayesian' with `nonparametric',
but indirectly acknowledged that the mathematical details
in such constructions would have to be more complicated.

\subsection{From the start to the present}

Independently of and concurrently with Lindley's review,
what may be considered to be the historical start 
of Bayesian nonparametrics took place in California.
The 1960ies had been a period of vigorous methodological 
research into various nonparametric directions. David Blackwell,
among the prominent members of the statistics department
at Berkeley (and, arguably, belonging to the Bayesian
minority there), suggested to his colleagues that 
there ought to be Bayesian parallels to problems and solutions,
for some of these nonparametric situations. These
conversations led to two noteworthy developments,
both important in their own rights and for what
followed. These were 
(i) a 1970 U.C.L.A.~technical report termed 
`A Bayesian analysis of some nonparametric problems',
by T.S.~Ferguson; and 
(ii) a 1971 University of Berkeley technical report called
`Tailfree and neutral random probabilities 
and their posterior distributions',
by K.A.~Doksum. These led after review processes
to the two seminal papers Ferguson (1973) 
in {\sl Annals of Statistics}, where the Dirichlet
process is introduced, 
and Doksum (1974) in {\sl Annals of Probability},
featuring his neutral to the right processes 
(see Chapters 2 and 3 for descriptions, 
inter-connections and further developments
of these classes of priors). The neutral to the right processes 
are also foreshadowed in Doksum (1972). 
In this very first wave of genuine Bayesian nonparametrics
work, also Ferguson (1974) stands out, an invited
review paper for {\sl Annals of Statistics}. 
Here he gives early descriptions of and results for 
P\'olya trees, for example, and points to further
fruitful research problems. 

We ought also to mention that there were earlier 
contributions to constructions of random probability measures
and their probabilistic properties, 
such as Kraft and van Eeden (1964) and 
Dubins and Freedman (1966). More specific Bayesian
connections, including matters of consistency
and inconsistency, were made in Freedman (1963)
and Fabius (1964), involving also the important
notion of tailfree distributions. Similarly, a density
estimation method given in Good and Gaskins (1971)
may be seen to have a Bayesian nonparametric root,
involving an implied prior on the set of densities.
Nevertheless, to the extent that such finer historical distinctions
are of interest, we would identify the start of
Bayesian nonparametrics with the work described
above by Ferguson and Doksum. 

These early papers provided significant stimulus
for many further developments, including research 
on various probabilistic properties of these new
prior and posterior processes (probability measures
on spaces of functions), procedures for density
estimation based on mixtures of Dirichlet processes,
applications to survival analysis (with suitable
priors on the random survivor functions, 
or cumulative hazard functions, and with methodology 
developed to handle censoring), a more flexible
machinery for P\'olya trees and their cousins, etc.
We point to Chapters 2 and 3 for further information, 
rather than detailing these developments here. 

The emphasis in this early round of new papers
was perhaps simply on the construction of new
prior measures, for an increasing range of 
natural statistical models and problems, along
with sufficiently clear results on how to 
characterise the consequent posterior distributions.
Some of these developments were momentarily
hampered or even stopped by the sheer 
computational complexity associated with 
handling the posterior distributions; sometimes
exact results could be written down and proved
mathematically, but algorithms could not always 
be constructed to evaluate these expressions. 
The situation improved around 1990, when 
simulation schemes of the MCMC variety 
became more widely known and implementable,
at around the time when statisticians suddenly
had real and easily programmable computers 
in their offices (the MCMC methods had in principle
been known to the statistics community since 
around 1970, but it took two decades for the
methods to become widely and flexibly used;
see e.g.~Gelfand and Smith, 1990). The MCMC methods
were at the outset constructed for classes
of finite-parameter problems, but it became
apparent that their use could be extended to
solve problems also in Bayesian nonparametrics. 

Another direction of research, in addition to the 
purely constructive and computational sides of 
the problems, is that of performance: how do 
the posterior distributions behave, in particular
when the sample size increases, and are the 
implicit limits related to those reached in 
the frequentist camp? Some of these questions
first surfaced in Diaconis and Freedman (1986a, 1986b),
where situations were exhibited in which 
the Bayesian machine yielded asymptotically 
inconsistent answers; cf.~also the many discussion
contributions to these two papers. 
This and similar research made it clearer to 
researchers in the field that even though asymptotics
typically lead to various mathematical statements 
of the comforting type 
`different Bayesians agree among themselves,
and also with the frequentist, as the sample 
size tends to infinity', for {\it finite-dimensional} problems, 
results are rather more complicated in 
infinite-dimensional spaces; 
cf.~Chapters 1 and 2 in this book
and comments already made in Section 1.1.

\subsection{Applications}

The subsection above dealt in essence with
theoretical developments. 
A reader sampling his or her way through the literarure
briefly surveyed there will make the anthropological
observation that articles written say after 2000
have a different look to them than those written 
say around 1980. This is partly reflecting a broader trend, 
with a transition of sorts that has moved
the primary emphases of statistics from the
more mathematically oriented articles to those
nearer to actual applications -- there are 
fewer sigma-algebras and less measure theoretic 
language, and more on motivation, algorithms, 
problem-solving and illustrations. 

The history of applications of Bayesian nonparametrics
is perhaps a more complicated and less well-defined 
one than that of the theoretical counterpart.
For natural reasons, including the general difficulty
of transforming mathematics to efficient algorithms
and the lack of good computers in the beginning of 
the nonparametric Bayes adventures, applications 
simply lagged behind. Ferguson's (1973, 1974) 
seminal papers are incidentally noteworthy also
since they spell out interesting and non-trivial
applications, e.g.~to adaptive investment models
and to adaptive sampling with recall, 
though without data illustrations.  
As indicated above, the first
broad theoretical foundations stem from the early
1970ies, while the first note-worthy real-data applications, 
perhaps primarily in the areas of survival analysis
and biostatistics, started to emerge in the early
1990ies (see e.g.~the book by Dey, M\"uller and Sinha, 1998).
At the same time various rapidly growing application areas
emerged inside machine learning (pattern recognition,
bioinformatics, language processing, search engines;
cf.~Chapter 7). More information and further pointers
to actual application areas for Bayesian nonparametrics
may be found by browsing the programmes for 
the Isaac Newton Institute workshop 2007
({\tt www.newton.ac.uk/programmes/BNR/index.html})
and that of the Carlo Alberto Programme in Bayesian
Nonparametrics 2009
({\tt bnpprogramme.carloalberto.org/index.html}). 

\subsection{Where does this book fit in the broader picture?}

We end this section by pointing to a short and
annotated list of books and articles 
in the literature that provide overviews of 
Bayesian nonparametrics (necessarily with 
different angles and emphases). 
The first and very early one of these is Ferguson (1974), 
mentioned above. 
Dey, M\"uller and Sinha (1998) is an edited
collection of papers, with an emphasis
on more practical concerns, and in particular
containing various papers dealing with survival analysis.
The book Ibrahim, Chen and Sinha (2001) gives 
a comprehensive treatment of the by then
more prominently practical methods
of nonparametric Bayes pertaining to survival analysis.
Walker, Damien, Laud and Smith (1999)
is a read discussion paper for the Royal Statistical Society,
exploring among other issues that of more 
flexible methods for P\'olya trees. 
Hjort (2003) is a later discussion paper, 
reviewing various topics and applications, 
pointing to research problems, and making connections 
to the broad `Highly Structured Stochastic Systems' theme
that is the title of the book in question.
Similarly M\"uller and Quintana (2004) 
provides another review of established
results and some evolving research areas. 
Ghosh and Ramamoorthi (2003) is an important
and quite detailed, mathematically oriented book 
on Bayesian nonparametrics, with focus 
on precise probabilistic properties of priors
and posteriors, including that of posterior
consistency (cf.~Chapters 1 and 2 of this book). 
Lee (2004) is a slim and elegant book dealing 
with neural networks via tools from Bayesian nonparametrics. 


\section{Further topics} 
\label{section:furthertopics}

Where would you want to go next (after having worked 
with this book)? The purpose of the present section 
is to rather briefly point to some of the research directions 
inside Bayesian nonparametrics that somehow lie outside 
the natural boundaries of the present book. 


\smallskip
{\it Gaussian processes:} 
Gaussian processes have an important role in several
branches of probability theory and statistics, 
also for problems related to Bayesian nonparametrics.
An illustration could be of regression data 
$(x_i,y_i)$ where $y_i$ is modelled as $m(x_i)+\eps_i$,
with say Gaussian i.i.d.~noise terms. If the unknown
$m(\cdot)$ function is modelled as a Gaussian process
with a known covariance function, 
then the posterior is another Gaussian process,
and Bayesian inference may proceed. 
There are many extensions of this simple scenario,
yielding Bayesian nonparametric solutions to 
different problems, ranging from prediction 
in spatial and spatial-temporal models 
(see e.g.~Gelfand, Guindani and Petrone, 2008)
to machine learning (cf.~Rasmussen and Williams, 2006).
Gaussian process models are also a popular choice 
for inference with output from computer simulation experiments; 
see e.g.~Oakley and O'Hagan (2002) and references there.
An extensive annotated bibliography of Gaussian process literature,
including links to public domain software, is
available at {\tt www.gaussianprocess.org/}.
Regression and classification methods using such processes
are reviewed in Neal (1999). 
Extensions to treed Gaussian processes is developed 
in Gramacy (2007) and Gramacy and Lee (2008).

\smallskip
{\it Spatial statistics:}
We touched on spatial modelling in connection
with the Gaussian processes above, and indeed many such models
may be handled, with the appropriate care, as 
long as the prior processes involved have 
covariance functions determined by a low number
of parameters. The situation is more complicated
when one wishes to place nonparametric priors 
also on the covariance functions; 
cf.~some comments in Chapter 4. 


\smallskip
{\it Neural networks:} 
There are by necessity several versions of `neural networks',
and some of these have reasonably clear Bayesian interpretations,
and a subset of these is amenable to nonparametric 
variations. See Lee (2004) for a lucid overview,
and e.g.~Holmes and Mallick (2000) for a particular application.


\smallskip
{\it $p>>n$ problems:}
A steadily increasing range of statistical problems
involve the `$p>>n$' syndrome, that there are much more
covariates (and hence unknown regression coefficients)
than individuals. Thus ordinary methods do not work,
and alternatives must be devised. Various methods
have been derived from frequentist perspectives,
but there is clear scope for developing Bayesian
techniques. The popular Lasso method of Tibshirani (1996)
may in fact be given a Bayesian interpretation,
as the posterior mode solution (the Bayes decision
under a sharp 0--1 loss function) 
with a prior for the large number of unknown
regression coefficients being that of independent
double exponentials with the same spread. 
Various extensions have been worked with, 
some also from this implied or explicit
Bayesian nonparametric perspective. 


\smallskip
{\it Model selection and model averaging:} 
Some problems in statistics are attacked by working 
out the ostensibly best method for each of a list
of candidate models, and then either select the 
tentatively best one, via some model selection criterion,
or average over a subset of the best looking ones. 
When the list of candidate models becomes large,
as it easily does, the problems take on nonparametric
Bayesian shapes; see e.g.~Claeskens and Hjort (2008, Ch.~7).
Further methodology needs to be developed for 
both the practical and theoretical side.

\smallskip
{\it Classification and regression trees:}
A powerful and flexible methodology for building regression
or classifiers via trees, with perhaps a binary option
at each node of the tree, was first developed 
in the CART system of Breiman, Friedman, Olshen and Stone (1984). 
Several attempts have been made at making Bayesian versions
of such schemes, involving priors on large families 
of growing and pruned trees. Their performance has been 
demonstrated to be excellent in several classes of problems; 
see e.g.~Chipman, George and McCulloch (2007). 
See in this connection also Neal (1999) mentioned above.

\smallskip
{\it Performance:} 
There are quite a few journal papers dealing with issues
of performance, comparisons between posterior distributions
arising from different priors, etc.; for some references
in that direction, see Chapters~1 and 2. 

\section{Computation and software} 
\label{section:software}

A critical issue for the practical use of nonparametric 
Bayesian prior models is the availability of efficient 
algorithms to implement posterior inference.
Recalling the earlier definition of nonparametric Bayesian 
models as probability models on big parameter spaces 
this might seem a serious challenge at first thought.  
But we run into some good luck. For many popular models 
it is possible to analytically marginalise with respect to 
some of the infinite-dimensional random quantities, 
leaving a probability model on some lower-dimensional 
manageable space. For example, under Gaussian process
priors the joint probability model for the realisation 
at any finite number of locations is simply 
a multivariate normal distribution. Similarly, 
various analysis schemes for survival and event history models 
feature posterior simulation of Beta processes (Hjort, 1990),
which may be accomplished by simulating and 
then adding independent Beta distributed increments
over many small intervals. 
Under the popular Dirichlet process mixture of normals model 
for density estimation the joint distribution of the observed
data can be characterised as a probability model 
on the partition of the observed data points 
and independent priors for a few cluster specific parameters. 
Also, under a P\'olya tree prior, 
or under quantile pyramids type priors (cf.~Hjort and Walker, 2009), 
posterior predictive inference can be implemented 
considering only finitely many levels of the nested 
partition sequence. 

Increased availability of public domain software 
for nonparametric Bayesian models greatly simplifies 
the practical use of nonparametric Bayesian models 
for data analysis. The perhaps most widely used
software is the R package {\sl DPpackage} (Jara, 2007,
exploiting the R platform of the R Development Core Team, 2006).
Functions in the package implement inference for 
Dirichlet process mixture density estimation, 
P\'olya tree priors for density estimation, 
density estimation using \Bernstein--Dirichlet priors, 
nonparametric random effects models,
including generalised linear models, 
semiparametric item-response type models, 
nonparametric survival models, inference for ROC 
(relative operating characteristic) curves
and several functions for families of dependent 
random probability models. 
See Chapter 6 for some illustrations. 
The availability of validated software like DPpackage 
will greatly accelerate the move of nonparametric Bayesian 
inference into the mainstream statistical literature.

\section{Challenges and future developments} 
\label{section:quovadem}

Where are we going, after all of this? 
A famous statistical prediction
is that `the twenty-first century will be Bayesian'.
This originated with Lindley's preface to
the English edition of de Finetti (1974),
and has since been repeated with different modifications
and different degrees of boldness by various
observers of and partakers in the principles and practice
of statistics; thus the {\sl Statistica Sinica}
journal devoted a full issue (2007, no.~2)
to this anticipation of the Bayesian century, for example.
The present book may be seen as yet another 
voice in this chorus, promising an increased
frequency of nonparametric versions of Bayesian methods.
Along with implications of certain basic principles, 
involving the guarantee of uncovering each possible 
truth with enough data (not only those truths that
are associated with parametric models), then,
in combination with the increasing versatility
and convenience of streamlined software, 
the century ahead looks decidedly both 
Bayesian and nonparametric. 

There are of course several challenges, 
associated with problems that are not yet solved 
in a sufficiently good manner, or that are perhaps
not worked with yet at the required level 
of seriousness. We shall here be bold enough 
to point to some of these.

Efron (2003) argues that the brightest statistical
future may be reserved for {\it empirical Bayes} methods, 
as tentatively opposed to pure Bayes methodology
that Lindley and others envisage. 
This points to the identifiable stream of Bayesian nonparametrics 
work that is associated with a careful setting
and fine-tuning of all the algorithmic parameters
involved in a given type of construction -- the 
parameters involved in a Dirichlet or Beta process,
or in an application of quantile pyramids modelling,
etc. A subset of such problems may be attacked
via empirical Bayes strategies (estimating these
hyper parameters via current or previously
available data) or by playing the Bayesian card
at a yet higher and more complicated level,
i.e.~via background priors for these hyper parameters. 

Another stream of work than may be surfacing is
that associated with replacing difficult and 
slow-converging MCMC type algorithms with quicker,
accurate approximations. Running MCMC in high dimensions,
as for several methods associated with models dealt
with in this book, is often fraught with difficulties 
related to convergence diagnostics etc. Inventing
methods that somehow sidestep the need for MCMCs
is therefore a useful endeavour. For good attempts
in that direction, for at least some useful and
broad classes of models, see Skaug and Fournier (2006)
and Rue, Martino and Chopin (2009). 

Gelman (2008), along with discussants, consider various
important objections to the theory and applications 
of Bayesian analysis; this is worthwhile reading
also since the writers in question belong 
to the Bayesian camp themselves. The themes they
point to, chiefly in a framework of parametric Bayes, 
are a fortiori valid for nonparametric Bayes. 

In Section 2 we pointed to the `two cultures' of 
modern statistics, associated respectively 
with the close interpretation of model parameters
and with automated black boxes. There are yet 
further schools or cultures, and an apparent
growth area is that broadly associated with 
{\it causality}. There are difficult aspects of 
theories of statistical causality, both conceptually
and model-wise, but the resulting methods see
steadily more application in e.g.~biomedicine,
see e.g.~Aalen and Frigessi (2007),
Aalen, Borgan and Gjessing (2008, Ch.~9)
and Pearl (2009). 
We predict that Bayesian nonparametrics will play
a more important role in such directions. 



\medskip
{\bf Acknowledgements.}
The authors are grateful to the Isaac Newton Institute
for Mathematical Sciences for making it possible for them 
to organise a broadly scoped programme on 
nonparametric Bayesian methods during August 2007.
The efforts and professional skills of the INI 
were particularly valuable regarding the international 
workshop that was held inside this programme,
with more than a hundred participants. 
They also thank Igor Pr\"unster for his many helpful
efforts and contributions in connection with the 
INI programme and the forum tutorial lectures. 

The authors also gratefully acknowledge support
and research environments conducive to their
researches in their home institutions 
(Department of Mathematics at the University of Oslo,
Department of Statistics at Oxford University, 
Department of Biostatistics at the University of Texas
M.D.~Anderson Cancer Center, and 
Institute of Mathematics, Statistics and Actuarial Science
at the University of Kent, respectively).
They are finally indebted to Diana Gilloly at 
Cambridge University Press for her consistently
constructive advice and for displaying 
the right amount of impatience.

\def\annals{Annals of Statistics}
\def\annprob{Annals of Probability}
\def\jasa{Journal of the American Statistical Association}
\def\sjs{Scandinavian Journal of Statistics}
\def\statprob{Statistics and Probability Letters}
\def\jrss{Journal of the Royal Statistical Society Series}
\def\statsci{Statistical Science}
\def\statmedres{Statistical Methods of Medical Research}
\def\adv{Advances in Applied Probability}
\def\compstat{Computational Statistics and Data Analysis}
\def\annmathstat{Annals of Mathematical Statistics}

\end{document}